\documentstyle[prl,aps,psfig,floats,times]{revtex}

\draft 
\begin{document}
\twocolumn[\hsize\textwidth\columnwidth\hsize\csname	%
@twocolumnfalse\endcsname				%
\title{New Relativistic Effects in the Dynamics of
Nonlinear Hydrodynamical Waves}

\author{Luciano Rezzolla$^{1,2}$ and Olindo Zanotti$^1$}

\address{$^1$SISSA, International School for Advanced Studies,
		Trieste, Via Beirut 2--4, 34014 Trieste, Italy}
\address{$^2$INFN, Department of Physics, University
		of Trieste, Via Valerio 2, 34127 Trieste, Italy} 

\date{\today}

\maketitle

\begin{abstract} 
In Newtonian and relativistic hydrodynamics the Riemann
problem consists of calculating the evolution of a fluid
which is initially characterized by two states having
different values of uniform rest-mass density, pressure
and velocity. When the fluid is allowed to relax, one of
three possible wave-patterns is produced, corresponding
to the propagation in opposite directions of two
nonlinear hydrodynamical waves. New effects emerge in a
special relativistic Riemann problem when velocities
tangential to the initial discontinuity surface are
present. We show that a smooth transition from one
wave-pattern to another can be produced by varying the
initial tangential velocities while otherwise maintaining
the initial states unmodified. These special relativistic
effects are produced by the coupling through the
relativistic Lorentz factors and do not have a Newtonian
counterpart.
\end{abstract}


\vskip2pc]


	More than a hundred years ago, Riemann addressed
the problem of the one-dimensional time evolution of a
perfect fluid which, at some given initial time $t=0$, is
composed of two adjacent states characterized by
different values of uniform velocity, pressure and
density~\cite{CF76}.  Since then this problem has been
referred to as the ``Riemann problem'' and is the
prototype of the initial value problem for hyperbolic
systems of partial differential equations with
discontinuous initial conditions. The conclusion reached
by Riemann in Newtonian hydrodynamics is that the
one-dimensional flow that develops when the barrier
separating the initial ``left'' (or 1) and ``right'' (or
2) states is removed, will allow for four different and
distinct solutions. All of the solutions are composed of
nonlinear waves, in the form of either shock waves or
rarefaction waves, that propagate in opposite directions
and join the two unperturbed left and right states. As a
result, three different ``wave-patterns'' can be produced
corresponding to two shock waves ($2S$), a shock and
rarefaction wave ($SR$) and two rarefaction waves ($2R$),
respectively.  Schematically, the fluid solution for any
$t>0$ can therefore be represented as
\begin{equation}
\label{general}
1\;{\cal W}_\leftarrow\; 3\; {\cal C}\;
	3'\;{\cal W}_\rightarrow \; 2	\ ,
\end{equation}
where ${\cal W}$ denotes a shock or a rarefaction wave
that propagates towards the left $(\leftarrow)$ or
towards the right $(\rightarrow)$ of the initial
discontinuity, $1$ and $2$ are the known initial states
with ``state-vectors'' consisting of fixed and
independent values of the pressure, rest-mass density and
velocity ${\mathbf U}_{1,2}\equiv(p,\rho,
v)_{1,2}$. Similarly, $3$ and $3'$ represent the new
hydrodynamic states that form behind the two waves
propagating in opposite directions. The two nonlinear
waves ${\cal W}_\leftarrow$ and ${\cal W}_\rightarrow$
are separated by a contact discontinuity ${\cal C}$
across which the rest-mass density is discontinuous, but
not the velocity nor the pressure, i.e.
$\rho_3\ne\rho_{3'},\; p_3=p_{3'},\; v_3=v_{3'}$. These
three quantities fully describe the states of the fluid
in the flow regions $3$ and $3'$ behind the nonlinear
waves. The formal solution of the Riemann problem
consists then in determining the velocity, pressure and
rest-mass density in the new states $3$ and $3'$ as well
as calculating, at any time $t>0$, the positions of the
waves separating the four states.

	The solution of the Riemann problem attracted a
wider interest when it was realized that its numerical
solution could be implemented as building blocks in the
construction of numerical Godunov-type methods for the
accurate solution of the hydrodynamical
equations~\cite{G59CW84}. In such methods, the
computational domain is discretized and each interface
between two adjacent grid-zones is used to construct the
initial left and right states of a ``local'' Riemann
problem. The evolution of the hydrodynamical equations is
then obtained through the solution of the sequence of
local Riemann problems set up at the interfaces between
successive grid-zones and basically consists of
determining the fluid pressure in regions $3$ and $3'$
(see ref.~\cite{MM99} for a recent review).

	The extension of the one-dimensional Riemann
problem to relativistic fluid velocities has been
performed by~\cite{T86} for the ``shock-tube'' problem
and by~\cite{MM94} for the general problem. Besides the
obvious additional complications introduced by special
relativity, the {\it one-dimensional} Riemann problem in
relativistic hydrodynamics {\it does not} show
qualitative differences from its Newtonian
counterpart. In a recent paper~\cite{RZ01}, we have
introduced a new approach to the solution of an exact
relativistic one-dimensional Riemann problem which
focuses on the relativistic invariant expression for the
relative velocity between the two unperturbed initial
states $(v^x_{12})_0 \equiv [(v^x_1 - v^x_2)/(1 -
v^x_1v^x_2)]_0$, where we have assumed that the
discontinuity is initially located on an $x={\rm const.}$
surface and where $v^x$ represents the component of the
4-velocity normal to this surface. This new approach
introduces great simplifications in the logical
formulation of the Riemann problem and has been essential
in revealing these new relativistic effects. For this
reason we will briefly review it in what follows.

	By construction, $(v^x_{12})_0$ measures the
relativistic jump of the fluid velocity normal to the
discontinuity surface. Within this new approach, the
solution to the relativistic Riemann problem is found
after the pressure in the region between the two
nonlinear waves has been calculated as the root of the
nonlinear equation~\cite{RZ01}
\begin{equation}
\label{our}
v^x_{12}(p_3) - (v^x_{12})_0 = 0 \ .
\end{equation}
The notation used in equation (\ref{our}) needs an
additional comment. It is should be emphasized, in fact,
that $p_3$ itself depends on the thermodynamical
properties of the initial states so that $v^x_{12} =
v^x_{12}(p_1, \rho_1, p_2, \rho_2)$. Note also that
$v^x_{12}(p_3)$ does not depend on the value of the
initial tangential velocities and has a functional form
that is different for each of the three possible
wave-patterns that might result from the decay of the
initial discontinuity. In ref.~\cite{RZ01} it was then
shown that the wave-pattern produced in a one-dimensional
relativistic Riemann problem by the decay of the
discontinuity can be {\it predicted} from knowledge of
the initial data. The possibility of predicting the
wave-pattern from the initial conditions applies also to
Newtonian hydrodynamics~\cite{LL87} and represents an
important advantage since it allows one to deduce in
advance which set of equations to use for the solution of
the Riemann problem. From a mathematical point of view,
the validity of the approach discussed in
ref.~\cite{RZ01} is based on the proof that the three
branches of the function $v^x_{12}=v^x_{12}(p_3)$,
corresponding to the three possible wave-patterns, are
monotonically increasing with $p_3$. Furthermore, it was
shown that the three different branches always join
smoothly through specific values of $v^x_{12}(p_3)$
denoted respectively as $({\tilde v^x}_{12})_{_{2S}}$ and
$({\tilde v^x}_{12})_{_{SR}}$. (In ref.~\cite{RZ01}
another limiting value, $({\tilde v^x}_{12})_{_{2R}}$,
was found but this is not relevant in the present
discussion and will not be considered here.)

	Exploiting these properties, an improved exact
Riemann solver can be built in which the comparison of
$(v^x_{12})_0$ with the relevant limiting values
$({\tilde v^x}_{12})_{_{2S}}$ and $({\tilde
v^x}_{12})_{_{SR}}$ constructed from the initial
conditions allows to determine, prior to solving equation
(\ref{our}), both the wave-pattern produced and the
functional form of $v^x_{12}(p_3)$ to be used in equation
(\ref{our})~\cite{RZ01}. In practice, this approach leads
to a straightforward numerical implementation and a more
efficient numerical algorithm with computational costs
reduced by up to 30\%.

	In generic multidimensional flows, the Riemann
problem can still be cast in terms of a planar
discontinuity surface across which a normal flow with
velocity $v^x$ takes place and is responsible for the
mass transfer. In general, however, the flow will also
have components tangential to the discontinuity surface,
which we will refer to as $v^t \equiv [(v^y)^2 +
(v^z)^2]^{1/2}$. The latter can also be discontinuous
across ${\cal C}$ possibly leading to a Kelvin-Helmoltz
instability there.  In relativistic regimes, the
tangential component cannot be removed by a single
Lorentz boost and is constrained to satisfy $|v^t| < [ 1
- (v^x)^2]^{1/2}$.

	A first study of a multidimensional relativistic
Riemann problem involving two strong rarefactions was
carried out in ref.~\cite{FK96} and has been extended to
a more general treatment in ref.~\cite{PMM00} where it
was pointed out that the tangential velocities in
multidimensional flows introduce differences between the
Newtonian and the relativistic Riemann problems. In
Newtonian hydrodynamics, in fact, the solution to the
Riemann problem does not depend on the tangential
component of the flow. In relativistic hydrodynamics, on
the other hand, the different regions of the flow are
coupled, both in the velocities and in the specific
enthalpy, through the Lorentz factors $W_{i}\equiv [1 -
(v^x_i)^2 - (v^t_i)^2]^{-1/2}$, where $i=1, 2$. This is
an important difference whose consequences are much
harder to assess.
\begin{figure}[htb]
\centerline{
\psfig{file=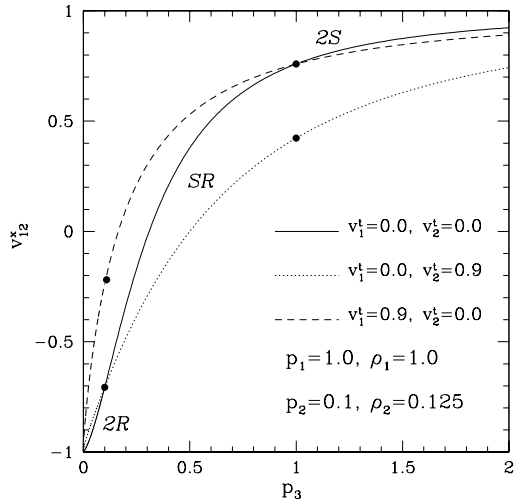,angle=0,width=7.0cm,height=7.0cm}
        }
\caption{Relative velocity as a function of the pressure
in region 3. The continuous line refers to the case of
zero tangential velocities while the dashed and dotted
ones refer to cases in which the tangential velocities
are nonzero (We use units in which $c=1$.). The filled
circles represent the joins between the solutions
consisting of two shock waves ($2S$), of one shock and
one rarefaction wave ($SR$) and of two rarefaction waves
($2R$).}
\label{fig1}
\end{figure}

	To discuss the {\it new} effects that emerge in a
{\it multidimensional} and {\it relativistic} Riemann
problem we have plotted in Fig.~\ref{fig1} the function
$v^x_{12}$ for different values of the pressure at the
contact discontinuity $p_3$. The curves shown refer to
the case of zero tangential velocities (continuous line)
and to the cases where the tangential velocities are
nonzero (dashed and dotted lines), respectively (Note
that the curves do not depend on the sign of $v^t$.). In
both cases the relative velocities are calculated for a
perfect fluid with polytropic equation of state
$p=k\rho^{\gamma}$, with a standard set of the initial
values of pressure and rest-mass density~\cite{S78},
i.e. $p_1=1.0, p_2=0.1,\; \rho_1=1.0, \rho_2=0.125$ (we
have here chosen $\gamma=5/3$). The filled dots in each
curve indicate the locations where the branches
representing the three wave-patterns ($2S, SR,$ and $2R$)
merge. The values of the relative normal-velocities where
this happens are just the values of $({\tilde
v^x}_{12})_{_{2S}}$ and $({\tilde v^x}_{12})_{_{SR}}$
mentioned above.

	A rapid examination of Fig.~\ref{fig1} is
sufficient to appreciate that the presence of tangential
velocities can introduce new qualitative differences in a
relativistic Riemann problem. When tangential velocities
are present, in fact, the relative normal-velocity
$v^x_{12}$ is a function of $p_3$ but, through the
coupling introduced by the Lorentz factors, also of
$v^t_1$ and $v^t_2$. This can introduce a {\it new
relativistic effect} and {\it shift the solution from one
wave-pattern to another}. To better appreciate this, let
us restrict attention to a simpler situation in which
only one of the two initial tangential velocities is
allowed to vary.
\begin{figure}[htb]
\centerline{
\psfig{file=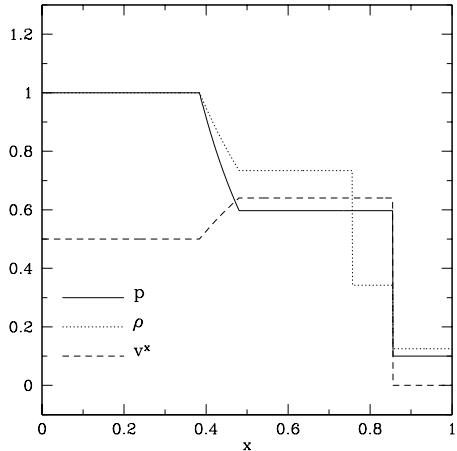,angle=0,width=6.5cm,height=6.5cm}
        }
\caption{Solution of the Riemann problem with initial
conditions: $p_1=1.0, p_2=0.1,\; \rho_1=1.0,
\rho_2=0.125$. The initial velocities are
$v^t_1=0.0=v^t_2$ and $v^x_1=0.5,\;v^x_2=0.0$.}
\label{fig2}
\end{figure}

\noindent Suppose now that the normal velocities are
chosen to be $v^x_1=0.5,\;v^x_2=0.0$, and that there are
no tangential velocities. In this case, $(v^x_{12})_0 =
0.5$ and Fig.~\ref{fig1} shows that the solution to the
Riemann problem falls in the $SR$ branch, hence producing
a wave-pattern consisting of a shock and a rarefaction
wave moving in opposite directions. This is presented in
Fig.~\ref{fig2} which shows the solution of the Riemann
problem at a time $t>0$ for the pressure, the rest-mass
density and the velocity.
\begin{figure}[htb]
\centerline{
\psfig{file=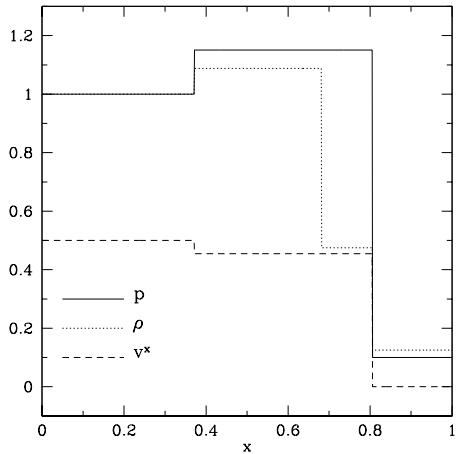,angle=0,width=6.5cm,height=6.5cm}
        }
\caption{As in Fig.~\ref{fig2} but with
$v^t_1=0.0$, $v^t_2=0.9$, $v^x_1=0.5$, $v^x_2=0.0$.}
\label{fig3}
\end{figure}
	However, if we now maintain the {\it same}
initial conditions but allow for nonzero tangential
velocities in state 2, Fig.~\ref{fig1} also shows that
the solution to the Riemann problem can fall in the $2S$
branch (cf. dotted line), hence producing a wave-pattern
consisting of two shock waves moving in opposite
directions. This is presented in Fig.~\ref{fig3} which
shows the solution of the same Riemann problem as in
Fig.~\ref{fig2} but with the initial tangential
velocities being $v^t_1=0.0$ and $v^t_2=0.9$. Note that
except for the tangential velocities, the solutions in
Figs.~\ref{fig2} and~\ref{fig3} have the same initial
state-vectors but different intermediate ones (i.e.
$p_3,\;\rho_3$, and $ v^x_3$).

	The Riemann problems shown in Figs.~\ref{fig2}
and~\ref{fig3} are only two possible examples but they
are useful ones since they show that through the coupling
among the different states introduced by the Lorentz
factors, a difference in the tangential velocities can
produce a smooth transition from one wave-pattern to
another while maintaining the initial states unmodified.
Interestingly, this transition does not need to always
produce a solution consisting of two shock waves.
\begin{figure}[htb]
\centerline{
\psfig{file=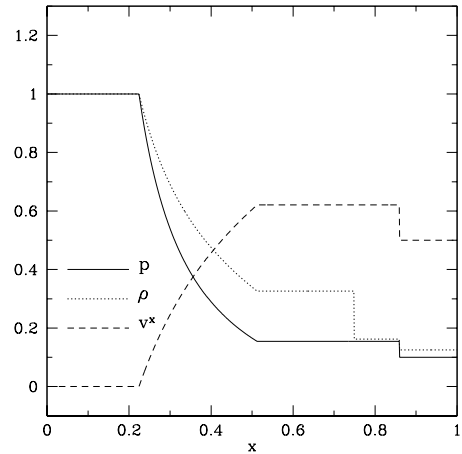,angle=0,width=6.5cm,height=6.5cm}
        }
\caption{As in Fig.~\ref{fig2} but with
$v^t_1=0.0=v^t_2$, $v^x_1=0.0$, $v^x_2=0.5$.}
\label{fig4}
\end{figure}

\noindent Suppose, in fact, that the normal velocities
are now chosen to be $v^x_1=0.0,\;v^x_2=0.5$. We can then
repeat the considerations made above and start examining
the wave-pattern produced when there are zero tangential
velocities. In this new setup, $(v^x_{12})_0 = -0.5$ and
Fig.~\ref{fig1} shows that the solution to the Riemann
problem still falls in the $SR$ branch (cf. dashed line),
with the corresponding solution at a time $t>0$ being
presented in Fig.~\ref{fig4}. Note that the wave-patterns
in Fig.~\ref{fig2} and~\ref{fig4} both consist of a shock
and a rarefaction wave, but have alternate initial normal
velocities.
\begin{figure}[htb]
\centerline{
\psfig{file=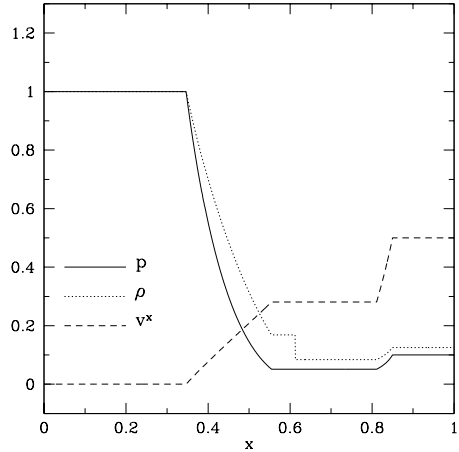,angle=0,width=6.5cm,height=6.5cm}
        }
\caption{As in Fig.~\ref{fig2} but with
$v^t_1=0.9$, $v^t_2=0.0$, $v^x_1=0.0$, $v^x_2=0.5$.}
\label{fig5}
\end{figure}

	When nonzero tangential velocities are now
considered in state 1, Fig.\ref{fig1} shows that
$(v^x_{12})_0$ can fall in the $2R$ branch, hence
producing a wave-pattern consisting of two rarefaction
waves moving in opposite directions. This is again shown
in Fig.~\ref{fig5} where we have chosen initial
tangential velocities $v^t_1=0.9$ and $v^t_2=0.0$. In
this case too and except for the tangential velocities,
the solutions in Figs.~\ref{fig4} and~\ref{fig5} have the
same initial state-vectors but different intermediate
ones.

	As mentioned above, the appearance of these new
relativistic effects is related to the behaviour of the
function $v^x_{12}$ for different values of the initial
tangential velocities and in particular to how the three
branches composing the curve change under variation of
$v^t_{1,2}$. As a result, the occurrence of these effects
can be recast into the study of the functional behaviour
of the values of $v^x_{12}$ delimiting the different
branches when the tangential velocities are varied within
the allowed ranges. A detailed mathematical treatment of
this problem will be presented in a separate
paper~\cite{RZ02} but it is sufficient to point out in
what follows the main results. Firstly, the dependence of
the limiting values for the $SR$ and $2S$ branches on the
tangential velocities can be expressed as
(cf. Fig.~\ref{fig1})
\begin{equation}
({\tilde v^x}_{12})_{_{SR}} = 
	({\tilde v^x}_{12})_{_{SR}}(v^t_1)\ ,
\hskip 1.0cm
({\tilde v^x}_{12})_{_{2S}} = 
	({\tilde v^x}_{12})_{_{2S}}(v^t_2) \ .
\end{equation}
Secondly, these values converge to zero when the
tangential velocities are allowed to reach their maximum
values, i.e.
\begin{equation}
\label{limits}
\lim_{W_1\rightarrow\infty} 
	({\tilde v^x}_{12})_{_{SR}}= 0 =
\lim_{W_2\rightarrow\infty} 
	({\tilde v^x}_{12})_{_{2S}} \ .  
\end{equation}
Expressions (\ref{limits}) indicate that for tangential
velocities assuming increasingly larger values, the $SR$
branch of the $v^x_{12}$ curve spans a progressively
smaller interval of relative normal-velocities. When the
tangential velocities reach their asymptotic values, the
$SR$ branch reduces to a point. As a result, a
wave-pattern consisting of a shock and a rarefaction wave
is generically disfavoured by increasing tangential
velocities.

	The effects discussed so far have a purely
special relativistic origin and might conflict with our
physical intuition. However, a behaviour similar to the
one reported here is found in the well-known relativistic
transverse-Doppler effect, in which the wavelength of a
photon received from a source moving at relativistic
speeds changes also if the source has a velocity
component orthogonal to the direction of emission of the
photon~\cite{R80}. In this case too, a Lorentz factor
including the transverse velocity is responsible for the
effect.

	A couple of comments are worth mention. Firstly,
there exists a set of initial conditions for which these
new relativistic effects will not occur. These initial
conditions are those of the classic ``shock-tube''
problem, in which $v^x_1=0=v^x_2$. In this case, in fact,
$(v^x_{12})_0 = 0$ and, because of the limits
(\ref{limits}), the solution of the Riemann problem will
be given by a wave-pattern consisting of a shock and a
rarefaction wave, independently of the values of the
tangential velocities. Secondly, while the stability of
shock fronts has been studied in the past
(see~\cite{AR86} and references therein), little is still
know about the stability properties of shock fronts
across which the tangential velocities are
discontinuous. It is plausible that a nonzero mass flux
makes these fronts stable with the respect to the
Kelvin-Helmoltz instability, but more work needs to be
done in the investigation of the stability of generic
shocks fronts.

	The results reported in this letter can be used
in astrophysical scenarios, such as those involving
relativistic jets or $\gamma-$ray bursts, in which
nonlinear hydrodynamical waves with large Lorentz factors
and complex multidimensional flows are
expected~\cite{rev}. Furthermore, both the approach to
the solution of the Riemann problem and the new
relativistic effects discussed here could be relevant in
the implementation of special relativistic exact Riemann
solvers in multidimensional codes solving the
hydrodynamical equations in curved
spacetimes~\cite{PMM00,Petal98}.

\medskip

	It is a pleasure to thank J.C. Miller and
J.A. Pons for useful discussions. Support for this
research has been provided by the MIUR and by the EU
Network Programme
(Research Training Network
Contract HPRN-CT-2000-00137).



\begin{references}

\bibitem{CF76}{R.~Courant and K.~O.~Friedrichs,}
	{\it Supersonic Flows and Shock Waves}, 
	Springer-Verlag, New York (1976)

\bibitem{G59CW84}{S.~K.~Godunov,} {\it Mat. Sb.}, {\bf 47},
	271 (1959); ~P.~Colella and P.~R.~Woodward, {\it
	J. Comput. Phys.}, {\bf 54}, 174 (1984)

\bibitem{MM99}{J.~M.~Mart\'{\i} and E.~M\"uller, }
	{\it Living Reviews}, {\bf 3} (1999)

\bibitem{T86}{K.~W.~Thompson,}
	{\it J. Fluid Mech.}, {\bf 171}, 365 (1986)

\bibitem{MM94}{J.~M.~Mart\'{\i} and E.~M\"uller, }
	{\it J. Fluid Mech.}, {\bf 258}, 317 (1994)

\bibitem{RZ01}{L.~Rezzolla and O.~Zanotti,}
	{\it J. Fluid Mech.} {\bf 449}, 395 (2001)

\bibitem{LL87}{L.~D.~Landau and E.~M.~Lifshitz,},
	{\it Fluid Mechanics (Second Edition)}, Pergamon Press
	(1987)

\bibitem{FK96}{ S.~A.~E.~G.~Falle and S.~S.~Komissarov},
	{\it Mon. Not. Royal Astron. Soc.}, {\bf 278}, 586 (1996)

\bibitem{PMM00}{J.~A.~Pons, J.~M.~Mart\'{\i} and E.~M\"uller,}
	{\it J. Fluid Mech. }{\bf 422}, 125 (2000)

\bibitem{S78}{G.~A.~Sod}, {\it J. Comp. Phys.}, {\bf
	27}, 1 (1978)

\bibitem{RZ02}{L.~Rezzolla and O.~Zanotti,}
	{\it In preparation} (2002)

\bibitem{R80}{W.~Rindler}, {\it Introduction to Special
	Relativity}, Clarendon Press, New York (1982)

\bibitem{AR86}{A.~M.~Anile and G.~Russo}, {\it Phys. of
	Fluids}, {\bf 29}, 2487 (1986); {\it ibidem} {\bf
	30}, 1045 (1987)

\bibitem{rev}{ R. D. Blandford}, {\it Prog. of
	Theor. Phys. Supp.}, {\bf 143}, in press (2002)
	{P.~Meszaros,}{\it Annu.~Rev.~Astron.~Astrophys.} {\bf
	40} in press (2002)

\bibitem{Petal98}{J.~A.~Pons, J.~A. Font, J.~M.~Mart\'{\i} J.~M.~
	Iba\~nez and J.~A.~ Miralles},
	{\it Astron. Astroph. }{\bf 339}, 638 (1998)

\end{references}
\end{document}